\documentclass[conference]{IEEEtran}
\IEEEoverridecommandlockouts
\usepackage{cite}
\usepackage{comment}
\usepackage{amsmath,amssymb,amsfonts}
\usepackage{algorithm}
\usepackage{algorithmic}
\usepackage{graphicx}
\usepackage{textcomp}
\usepackage{xcolor}
\usepackage{float}
\usepackage{multirow}
\usepackage{adjustbox}
\usepackage{cleveref}
\usepackage[caption=false]{subfig}
\def\BibTeX{{\rm B\kern-.05em{\sc i\kern-.025em b}\kern-.08em
    T\kern-.1667em\lower.7ex\hbox{E}\kern-.125emX}}
\begin{document}

% Membership Inference Attack: Batch-wise generated Attack Dataset

\title{MIA-BAD: An Approach for Enhancing Membership Inference Attack and its Mitigation with Federated Learning }

\author{
    \IEEEauthorblockN{
    Soumya Banerjee\IEEEauthorrefmark{1},
    Sandip Roy\IEEEauthorrefmark{1},
    Sayyed Farid Ahamed\IEEEauthorrefmark{1},
    Devin Quinn\IEEEauthorrefmark{2},  
    Marc Vucovich\IEEEauthorrefmark{2},
    }
    \IEEEauthorblockN{ 
    Dhruv Nandakumar\IEEEauthorrefmark{2},
    Kevin Choi\IEEEauthorrefmark{2}, 
    Abdul Rahman\IEEEauthorrefmark{2},
    Edward Bowen\IEEEauthorrefmark{2},
    Sachin Shetty\IEEEauthorrefmark{1}
    }
    \IEEEauthorblockA{
    \IEEEauthorrefmark{1}Virginia Modeling, Analysis and Simulation Center, Old Dominion University, Virginia, USA
    \\\{s1banerj, sroy, saham001, sshetty\}@odu.edu}
    \IEEEauthorblockA{
    \IEEEauthorrefmark{2}Deloitte \& Touche LLP
    \\\{devquinn, mvucovich, dnandakumar,kevchoi, abdulrahman, edbowen\}@deloitte.com}
}

\maketitle

\begin{abstract}
The membership inference attack (MIA) is a popular paradigm for compromising the privacy of a machine learning (ML) model. MIA exploits the natural inclination of ML models to overfit upon the training data. MIAs are trained to distinguish between training and testing prediction confidence to infer membership information.   
Federated Learning (FL) is a privacy-preserving ML paradigm that enables multiple clients to train a unified model without disclosing their private data.
%In this paper, we present MIA: Batch-wise generated Attack Dataset (MIA-BAD), a modification to the MIA approach. 
In this paper, we propose an enhanced Membership Inference Attack with the Batch-wise generated Attack Dataset (MIA-BAD), a modification to the MIA approach. 
We investigate that the MIA is more accurate when the attack dataset is generated batch-wise. This quantitatively decreases the attack dataset while qualitatively improving it.
We show how training an ML model through FL,  has some distinct advantages and investigate how the threat introduced with the proposed  MIA-BAD approach can be mitigated with FL approaches. 
Finally, we demonstrate the qualitative effects of the proposed MIA-BAD methodology by conducting extensive experiments with various target datasets, variable numbers of federated clients, and training batch sizes.
\end{abstract}


\section{Introduction}

In recent years, federated learning (FL) has emerged as a popular privacy-preserving machine learning (ML) paradigm, allowing multiple clients to collaboratively develop a unified and consolidated model while protecting their individual training data \cite{mcmahan1602federated}. Unlike typical centralized ML, FL does not need users to send their original data to the centralized server, which might compromise users' privacy and security. The basic procedure involves training localized models on individual client data, followed by the exchange of model updates among federated clients, and finally the building of a unified model that is shared by all clients \cite{vucovich2022anomaly}. Because collaborative learning doesn't involve sharing data, FL shows the potential to solve problems with data privacy and user privacy that are common in traditional centralized ML \cite{chen2021pois}.

Despite the fact that FL is designed to secure personal information \cite{ma2020safeguarding}, recent research has shown that FL models are vulnerable to attacks that leak critical training dataset information through source inference, model inversion, and reconstruction attacks \cite{hu2021source}. In this paper, we focus on membership inference attacks (MIAs), which are aimed against the FL model, and attempt to infer whether or not the target sample was included in the model's training data \cite{shokri2017membership}. MIAs that are successful may jeopardize the security of federated clients \cite{nasr2019comprehensive}. For example, knowing the target sample may disclose the victim's condition and treatment history if an FL model is trained on data from different medical institutions\cite{crowson2022systematic}. An attacker can launch a membership inference attack in a ``Black-Box" environment (Restrictive Knowledge of the model) by developing a binary attack model that, when fed the confidence score vector from the target model, returns the likelihood that the given data sample is part of the training dataset \cite{shokri2017membership}.

% \begin{figure}
%      \centering
%      \begin{subfigure}{\linewidth}
%          \centering
%          \includegraphics[width=0.82\linewidth]{Fig_1a.png}
%          \caption{Training an ML model centrally}
%      \end{subfigure}
%      \hfill
%      \begin{subfigure}{\linewidth}
%          \centering
%          \includegraphics[width=\linewidth]{Fig_1b.png}
%          \caption{Training an ML model over FL}
%      \end{subfigure}
%      \caption{Training an ML model, Centrally Vs FL}
%     \label{fig:CentrallyVsFL}
% \end{figure}

% \begin{figure}[htp]
% \centering
% \subfloat[]{%
%   \includegraphics[width=0.8\columnwidth]{Fig_1a.png}%
% }

% \subfloat[]{%
%   \includegraphics[clip,width=\columnwidth]{Fig_1b.png}%
% }

% \caption{Training an ML model, Centrally Vs FL}
% \label{fig:CentrallyVsFL}
% \end{figure}

\begin{figure}[h!]
    \centering   
    \fbox{
    \includegraphics[width=0.45\textwidth]{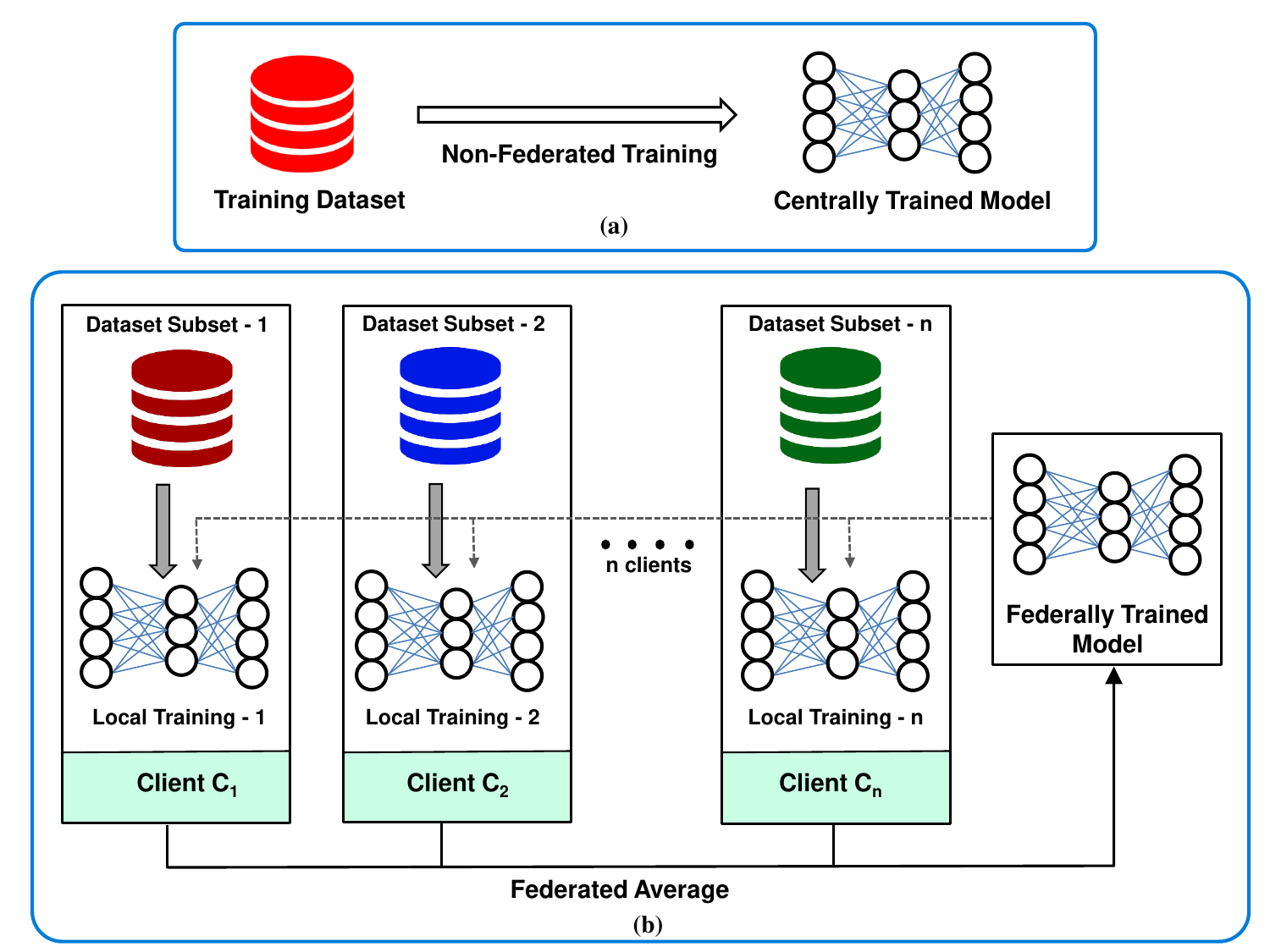}}
    \caption{Training procedure of an ML model, centrally vs FL. (a) Centrally or non-federated training of an ML model. (b) Training of an ML model in FL environment consisting of n clients. }
    \label{fig:CentrallyVsFL}
\end{figure}

Given a sample of data, MIA in FL detects its participation in the FL training process. A breach of privacy transpires when an adversary has any knowledge about the use of a certain data element for the purpose of training FL \cite{shokri2017membership}. For example, if the data includes device information, we may deduce if the device is participating in the FL training process, exposing the device's privacy \cite{ali2022federated}. 

%In this paper, we propose a modification to the MIA, that increases the attacker's advantage. 
In this paper, our aim is twofold. Initially, we show how an adversary can more accurately launch MIA attack into FL models for various datasets and then develop new insights for resisting such attacks in FL environments. We demonstrate that the FL paradigm while providing only limited protection to MIA, has a distinct advantage in reducing the potency of the proposed MIA-BAD approach.

The main contributions of this paper are as follows:

\begin{itemize}

    \item We demonstrate how federated training can reduce the potency of membership inference attacks and how the number of federated clients affects this result.
    
    % \item We propose a novel modification to the membership inference attacks paradigm, MIA-BAD, to increase the attacker's advantage.

    \item We propose a novel modification to the membership inference attacks paradigm, MIA-BAD, showing that the MIA is more accurate when the attack dataset is generated batch-wise.

    % \item Through detailed experiments we document the MIA-BAD approach and its qualitative effects.

    \item We demonstrate how the attacker's advantage of the proposed MIA-BAD can be minimized with FL. Through detailed experiments with different target datasets, and a varying number of clients and batch sizes, we document the the qualitative effects of the proposed MIA-BAD approach. 
\end{itemize}

The paper is organized as follows. In \cref{Priliminary}, we introduce the relevant theoretical background and present a brief literature survey. \Cref{threat} defines the threat model and the attacker's goals, knowledge, and capabilities. \Cref{overview} defines the framework for implementing the membership inference attack.  We propose the MAI-BAD approach in \cref{MIABAD}. The experimental results are provided and analyzed in \cref{eval}. Finally, we conclude our work and discuss future research scope in \cref{conclusion}.

\section{Preliminary}\label{Priliminary}

\subsection{Federated Learning}

FL is a decentralized ML training approach that allows multiple clients to collaboratively train a shared machine-learning model without accessing the local data of the clients \cite{yin2021comprehensive}. A conventional FL system consists of $n$ federated clients ($C_1, C_2, C_3,...,C_n$), in proximity with a central server denoted as $S$. The central server $S$ is responsible for coordinating the FL training process and through iterative training generates a converged global FL model $\mathcal{M}$.
The training process of the ML model $\mathcal{M}$ at the $r^{th}$ training round (where $r$ is an element of set $R$, representing the number of FL training rounds) is comprised of the following four steps:
\begin{itemize}
    \item Step 1. The centralized server denoted as $S$, distributes the current global FL model, denoted as $\mathcal{M}^r$, to the federated clients participating in the process, denoted as $C_i$ (where $i$ ranges from $1$ to $n$)
    \item  Step 2. In parallel, each client $C_i$ independently trains and improves the model $\mathcal{M}^r$, using its own dataset $D_i$. Once the local training process has been concluded, each client $C_i$ transmits the updated model parameters, denoted by $U_i^r$, to the central server. 
    \item Step 3. The central server $S$ accumulates the updated parameters $U^r = [U_1^r, U_2^r, \cdots, U_n^r]$ from each of the clients that are participating in the process. 
    \item Step 4. The central server $S$ then updates the global model $M^r$ by aggregating the collected parameter updates $U^r$. During the subsequent training iteration, the recently updated model $M^{r+1}$ will be distributed to federated clients. 
\end{itemize}

The training process in FL is executed iteratively by the central server and clients until a termination criterion is met. This criterion can be a maximum number of iterations or a threshold for model accuracy. Subsequently, the central server converges FL model $\mathcal{M}$, which is then distributed to each and every client of the system. \Cref{fig:CentrallyVsFL} contrasts the centralized and federated training of ML models.

In order to enhance privacy protection, recent advancements in the field of FL have incorporated various privacy protection mechanisms such as differential privacy\footnote{\textbf{Differential privacy} guarantees that an individual's data is modified or perturbed in a controlled manner to provide a privacy guarantee, ensuring that the overall statistical outcomes of computations are stable and independent of whether their specific information is included or altered.}\cite{geyer2017differentially} and secure aggregation\footnote{\textbf{Secure aggregation} ensures that data collected from multiple sources is combined using cryptographic techniques to maintain individual data privacy and confidentiality, enabling collaborative computations without revealing the raw data.}\cite{truex2019hybrid}. 
%\textcolor{red}{(Can we provide a brief summary of 1 sentence on differential privacy and secure aggregation for the reader? This can be footnotes... It will help with the background as this section is preliminary.)} 
Prior studies have primarily concentrated on two approaches for achieving differential privacy: centralized differential privacy, which relies on a central trusted party \cite{geyer2017differentially}, \cite{mcmahan2017learning}, and local differential privacy, where each user perturbs their updates randomly before transmitting them to an untrusted aggregator \cite{truex2019hybrid}, \cite{sun2020ldp}. While FL has been recognized as a potentially effective approach that prioritizes privacy, a limited number of recent studies have demonstrated the susceptibility of FL to MIAs \cite{nasr2019comprehensive}, \cite{lee2021digestive}. 
%\textcolor{red}{(Ok this last sentence is repeated. Do we want to restate this ?)}

\subsection{Membership Inference Attack}
Our work focuses on investigating the potential of membership privacy breaches through MIA and studying the attack's accuracy. In this section, we provide a brief overview of the background of existing works of MIAs as they pertain to ML models. An  attacker intends to ascertain the membership property of a target sample $x$, i.e., whether or not this sample was used to train the target model, by exploiting the prediction behavior of an ML model $\mathcal{T}_m$ (referred to as the target model). MIA may compromise the privacy of training data used in ML models, thereby introducing additional risks for the producers of such training data.

The MIA can be expressed in a more formal way as: 

\begin{equation}
 \mathcal{A}(\mathcal{T}_m,\Omega ,x) \rightarrow \textbf{In}\,/\,\textbf{Out} 
\end{equation}

where the attack is represented by $\mathcal{A}$, \textbf{In} means that $x$ is a member, \textbf{Out} means $x$ is not a member of training data $\mathcal{T}_m$. $\Omega$ refers to any additional knowledge about the target model and its training data that $\mathcal{A}$ can receive.

In 2017, Shokri et al. \cite{shokri2017membership} introduced an MIA algorithm for ML models, that assumes that the adversary has ``Restrictive Knowledge" access to the model\footnote{\textbf{Restrictive Knowledge} is achieved by an adversary for a black-box access or query access. It implies that the attacker can query the ML model to get a prediction on a given sample, but has no other access to the ML model and its weights.} and similar sample data as a priori knowledge. The experiments showed that the adversary can obtain significant information about the training data by attacking the model. In an FL environment, Truex et al. \cite{truex2019demystifying} gave insights that MIA may also result in the exposure of private membership information. Salem et al. \cite{salem2018ml} showed that the criteria for the execution of an MIA may be relaxed, which allows the attack to be carried out in a wider range of settings. 
%Melis et al. \cite{melis2019exploiting} claimed that in a collaborative learning environment, membership inference was more effective through gradient updates policy. 
Xu et al. provided a method for conspiring attackers to purposefully alter training data in order to enhance or decrease the weight of a dimension in the aggregation model \cite{xu2020information}. If the aggregation model's weights or parameters reflect a recognizable pattern that generates relevant signals, sensitive data may be exposed. 
\par \textbf{Synthesis: }Recently, there have been a few efforts to deploy various MIA variants in both centralized and federated ML environments. However, the effect of attack accuracy on different batch sizes of training to build the attack model for different numbers of clients in the FL has not been well studied. 

%\textcolor{red}{(Can we make 1-2 sentences that synthesize the literature above?)}

\section{Threat Model} \label{threat}

In this section, we discuss the basic threat model, where we explain the attacker's knowledge of the environment, the attacker's goal, and the capability \cite{shokri2017membership, salem2018ml, truex2019demystifying}.

\textbf{Adversary’s knowledge:} We consider the challenging situation in which the adversary can only acquire restricted knowledge about the target model and knowledge about the distribution of the target dataset. The training goals and model architecture are shared by all FL participants. As a result, this level of attacker knowledge is realistic. However, the adversary is unable to get any knowledge on the global training process (centralized or FL), or the distribution of the training data among the clients (if applicable).

\textbf{Attacker’s Goals:} For the ML model, the attacker guesses or infers data from the initial training set. Following training, the attacker develops an attack model that infers private data from query-level access to the target model. The attacker does not alter the parameters of the model and does not need any additional information \cite{shokri2017membership}.

\textbf{Attacker’s Capability:} The attacker is presumed to be an honest but curious user with query access to the target model, but cannot access its internal weights and gradients. The attacker uses this query access to implement the membership inference attack.

\begin{figure}[h!]
    \centering   
    %\fbox{
    \includegraphics[width=0.48\textwidth]{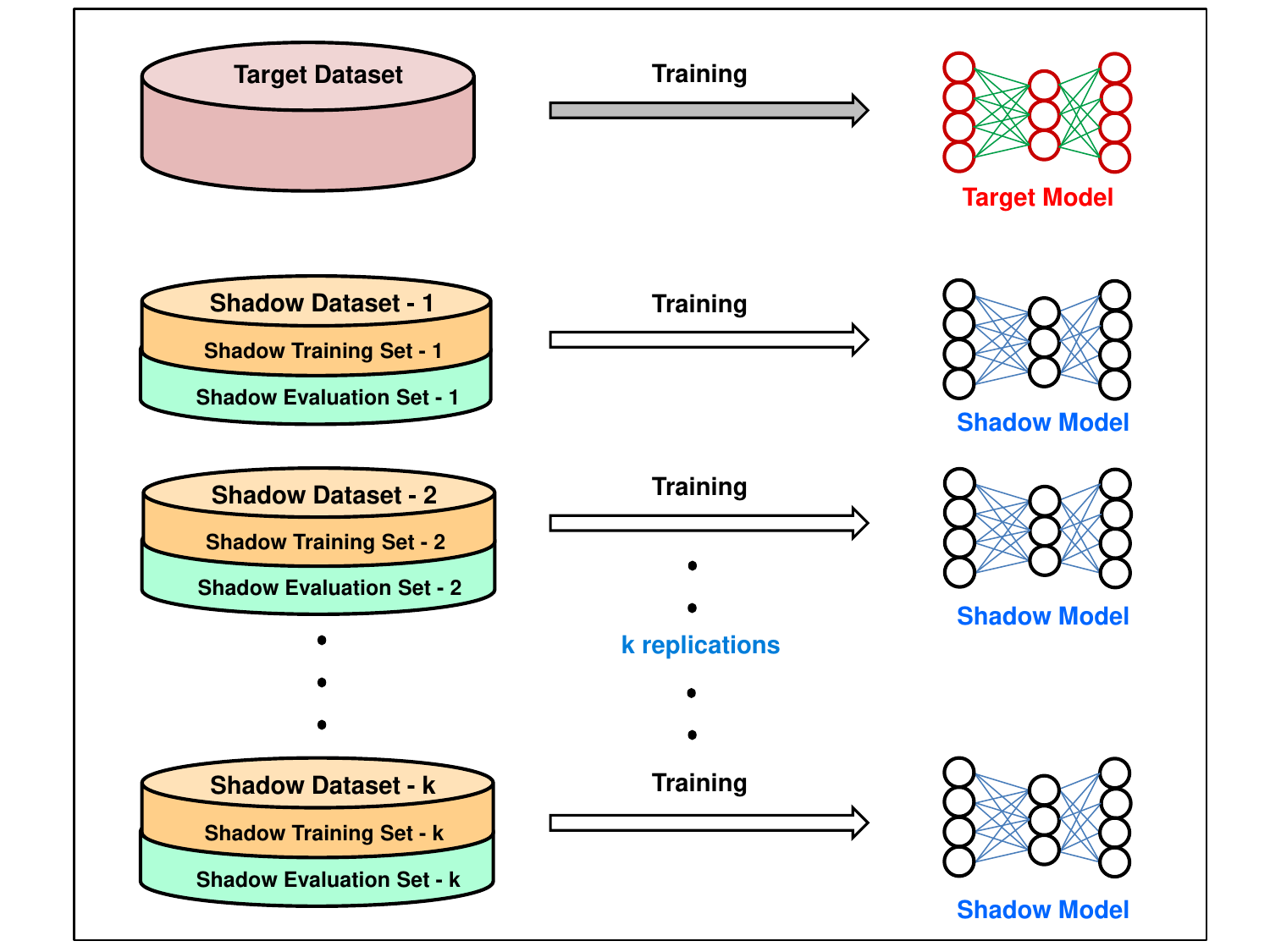}
    %}
    \caption{MIA's shadow model training technique. Shadow models are trained in the same manner as the target model using $k$ shadow-training and shadow-evaluation datasets.}
    \label{fig:shadow_training}
\end{figure}

\section{Framework Overview}\label{overview}
\subsection{Attack Overview}
Our study is focused on a framework that is predicated on the notion that, given training vs. non-training data, MIA may recognize the difference in an ML model's behavior. The inference attacker employs a two-step strategy. To begin, a collection of shadow models with behaviors comparable to the target model is built and trained. Then, the shadow models are utilized to generate an attack dataset for training the attack model. The trained attack model can infer whether a sample belongs to the training set of the target model data.

When the FL training procedure is completed, a global model is produced. This global model comprises the training outcomes of all participants' data that comes from many rounds of aggregation. To the adversary, this federally trained model is indistinguishable from a centrally trained model and thus makes no difference for MIA purposes.

The attacker trains the shadow models locally, each with behavior comparable to the global model. Having trained on the shadow model outcomes, the attack models may effectively infer the global model's ``in" and ``out" dataset.

\subsection{Shadow Model}
The concept of shadow model training involves using identically distributed data on the as similar as possible model architecture to train models that evaluate data similarly \cite{shokri2017membership}. The attacker utilizes knowledge about the target dataset's distribution to build multiple shadow datasets with similar styles and distributions. The attacker trains shadow models, with a similar architecture to the target model, on these shadow datasets. Once trained, the shadow model is used to produce training data for the attack model. \Cref{fig:shadow_training} summarizes how the shadow models are trained.

% \begin{figure}[h!]
%     \centering   

%     \includegraphics[width=0.48\textwidth]{Fig3-1.png}
%     \caption{Membership Inference Attack, attack dataset generation}
%     \label{bulid_attack_dataset}
% \end{figure}
% \subsection{Attack Model}

\begin{figure*}[!htb]
    \centering
	\fbox{\includegraphics[width=0.6\linewidth]{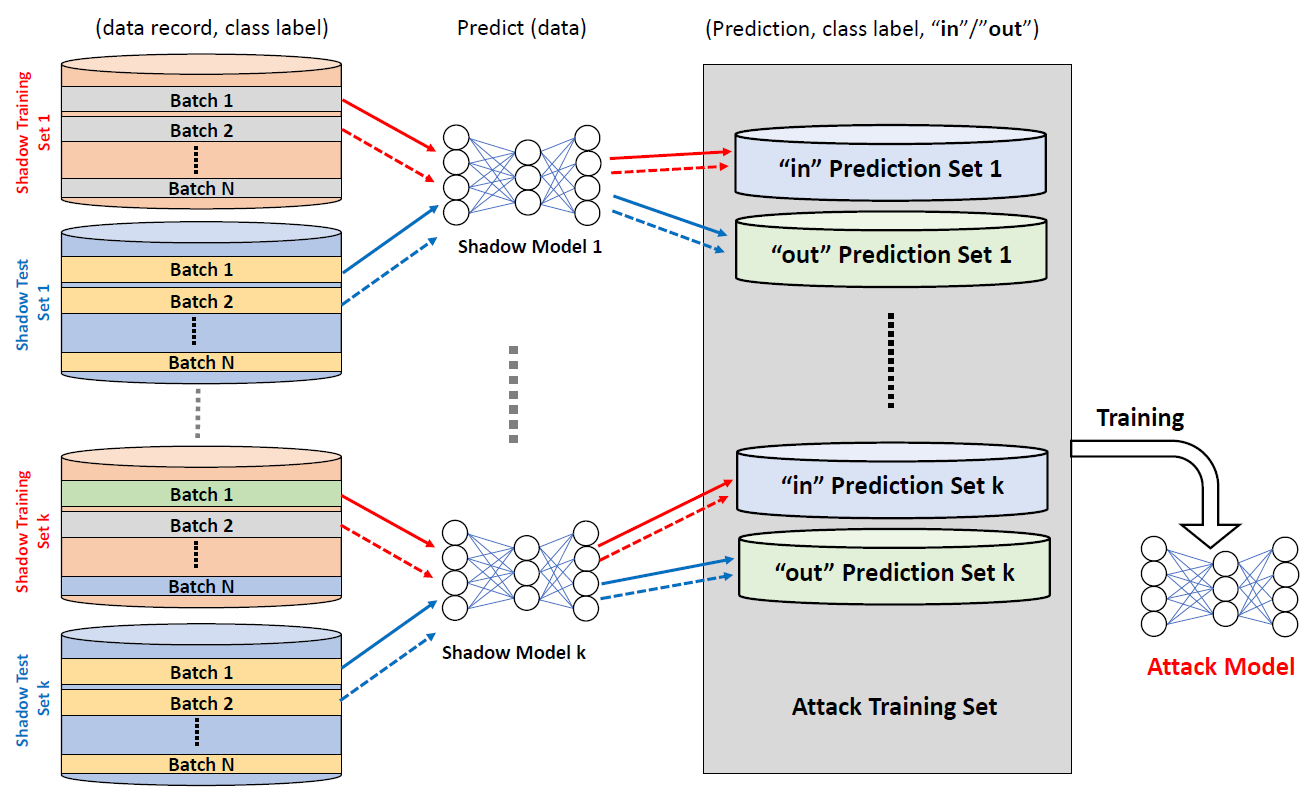}}
    \caption{Training process of the proposed MIA-BAD approach. The attack model is trained from batch-wise generation of the attack dataset.}
    \label{bulid_attack_dataset}
\end{figure*}

\subsection{Attack Model} \label{attack_model}
The attack model is a binary classifier that is trained 
%The attacker uses the attack model's binary classification property 
to predict if a given record is part of the model's training set or not. For each record in the shadow model's training dataset, the adversary creates a query, retrieves the output, marks the resultant vectors as ``in" and finally adds them to the attack model's training dataset. Similarly, the attacker queries and records output labeled as ``out" using a disjoint test dataset of the shadow model. The attacker now possesses a dataset comprising records, as well as the appropriate outputs of the shadow models and the corresponding in/out labels. The objective of the attack model is to infer the labels from the records and corresponding outputs.

% \begin{figure*}[!htb]
%     \centering
% 	\fbox{\includegraphics[width=0.6\linewidth]{fig3-2.png}}
%     \caption{Training process of the proposed MIA-BAD. The attack model is trained from batch-wise generation of the attack dataset.}
%     \label{bulid_attack_dataset}
% \end{figure*}

\section{The MIA-BAD Approach}\label{MIABAD}

In this section, we investigate the effect of modifications on how the attack model is trained.  Normally, as described in \cref{attack_model}, the attack dataset is built by evaluating on the shadow model to enable tabulation of the sample-wise loss in the attack dataset.   This approach produces a dataset equal to the size of all the shadow datasets combined. This gives us more than enough data to train the attack model.

However, it is a well-known phenomenon that ensembling improves performance \cite{dietterich2000ensemble}.  We generally 
%\textcolor{red}{we should avoid making blanket statements like this and should rephrase to 'generally' or 'commonly'} 
train ML models in batches because larger batch sizes can provide better model convergence \cite{keskar2016large}. In an MIA, the shadow models as well as the attack models are trained in batches, however, the attack dataset is built sample-wise to retain the size of the attack dataset.   

We propose, building the attack dataset by evaluating batch-wise with shadow models on the corresponding shadow dataset. This will significantly decrease the size of the attack dataset. Nevertheless, we hypothesized that the qualitative improvement through the implied ensembling would compensate for the quantitative loss, \cref{Res-C} backs our hypothesis. \Cref{bulid_attack_dataset} demonstrates the MIA-BAD approach for attack model training.

\subsection{The Overall Attack Paradigm}\label{attack_summary}
%In this section, we present the proposed MIA with a Batch-wise generated Attack Dataset (\textbf{MIA-BAD}) paradigm. \Cref{algo} provides the pseudo-code for implementing the attack.
In this section, we first present the basic steps of the proposed MIA-BAD approach. Next, then provide the algorithm (in pseudo-code format), in \Cref{algo}, for implementing the proposed attack.

\begin{enumerate}
    \item Define a master shadow dataset by sampling from a similar distribution as the target dataset.

    \item Sample $k$ overlapping but significantly different shadow dataset from the master shadow dataset. Split each shadow dataset into training and test subsets.

    \item Train the $k$ shadow models on the training subsets of its corresponding shadow dataset.

    \item For every shadow model, evaluate the entire shadow dataset (train or test, represented by seen and unseen) batch-wise, and record the batch-wise loss with the seen/unseen label to build the attack dataset. 

    \item Train a binary classifier on the attack dataset as the attack model. 
   \end{enumerate}

\begin{table*}[!htb]
	\caption{Performance of MIA-BAD on CIFAR10 dataset for different batch sizes.} \label{Tab2}
	%\resizebox{\linewidth}{!}{%
        %\normalsize
		\begin{center}
			\begin{tabular}{ l|c|cccccc}
				\hline
				Training Mode & Sample wise &        \multicolumn {6}{c}{Batch wise}        \\\cline{3-8}
				              &             & $8$   & $16$  & $32$  & $64$  & $128$ & $258$ \\ \hline
				Centrally     & 0.833       & 0.853 & 0.869 & 0.856 & 0.853 & 0.869 & 0.856 \\
				$2$ clients   & 0.721       & 0.751 & 0.755 & 0.757 & 0.753 & 0.753 & 0.751 \\
				$5$ clients   & 0.769       & 0.769 & 0.772 & 0.768 & 0.771 & 0.768 & 0.769 \\
				$10$ clients  & 0.786       & 0.765 & 0.765 & 0.796 & 0.753 & 0.769 & 0.786 \\ \hline
			\end{tabular}
		\end{center}
	%}
\end{table*}

\begin{algorithm}[!htb]

	\caption{Algorithm for MIA-BAD} \label{algo}
  \textbf{Input:}  Target dataset distribution $\Omega$, Shadow model definition $\mathcal{S}$, Shadow count $k$, Batch size $B$\\
  \textbf{Output:} Attack model $\mathcal{A}$   %$\mathcal{T}_m,\Omega ,x) \rightarrow$ in /\ out$
		\begin{algorithmic}[1]
                \STATE Set, $\mathcal{D}$ = [] \hspace{0.8cm}    $\vartriangleright$ \textit{Attack Dataset.}
                \STATE Initialize, $\mathcal{A}\leftarrow$ Binary Classifier.
                \FOR{$i = 1$ to $k$} 
                \STATE Initialize, $\mathcal{S}_i\leftarrow  \mathcal{S}$.
                \STATE Sample, $\omega_i\leftarrow  \Omega$.
                \STATE Train $\mathcal{S}_i$ with $\omega^{train}_i$

                \FOR{batch $b$ $\in$ $\omega_i$} 
                \STATE Set, size of $b$ as $B$.   
                \STATE Evaluate $y \leftarrow\mathrm{S}_i(b)$
                \IF{$b \in \omega^{train}_i$}
                    \STATE Update $\mathcal{D}\leftarrow \langle y, in\rangle$ 
                \ELSIF {$b \in \omega^{test}_i$}
                    \STATE Update $\mathcal{D}\leftarrow \langle y, out\rangle$
                \ENDIF                  
                \ENDFOR
                \ENDFOR
                \STATE Train $\mathcal{A}$ with $\mathcal{D}$
                \RETURN {$\mathcal{A}$}
   
		\end{algorithmic}
\end{algorithm}

\section{Performance Evaluation}\label{eval}
In this section, we first define our experimental setup, then we demonstrate the effect of MIA on federally trained ML as opposed to a centrally trained model.  Finally, we demonstrate the effect of batch-wise generation of the attack dataset. 
\subsection{Experimental Setup}
We define a centralized FL architecture that trains ML models over a wide range of datasets. The architecture distributes the datasets into $n$ disjoint subsets and sends them to $n$ clients. Each client trains a local copy of the model architecture on the locally available data. After $e$ epochs of local training the clients (securely) share the model weights with the orchestrating server which combines all the received model weights through a federated average \cite{mcmahan2017communication} to define the global model, which is sent back to the clients.  For the next round, the client completes $e$ epochs of local training on the received model and shares the updated weights with the server. The server conducts $r$ such rounds to arrive at the global model. For a baseline comparison, an ML model with the same architecture is trained for $e\times r$ epochs. 

For this work, we evaluated the efficientnet \cite{tan2019efficientnet} architecture on the CIFAR10 \cite{krizhevsky2009learning}, CIFAR100 \cite{krizhevsky2009learning}, MNIST \cite{lecun1998mnist}, and FashionMNIST \cite{xiao2017fashion} datasets.

% For our research, we consider the test set of the target data set as the unseen data.

\subsection{Membership Inference Attack on FL}
\begin{figure}[h!]
    \centering   

    \includegraphics[width=\linewidth]{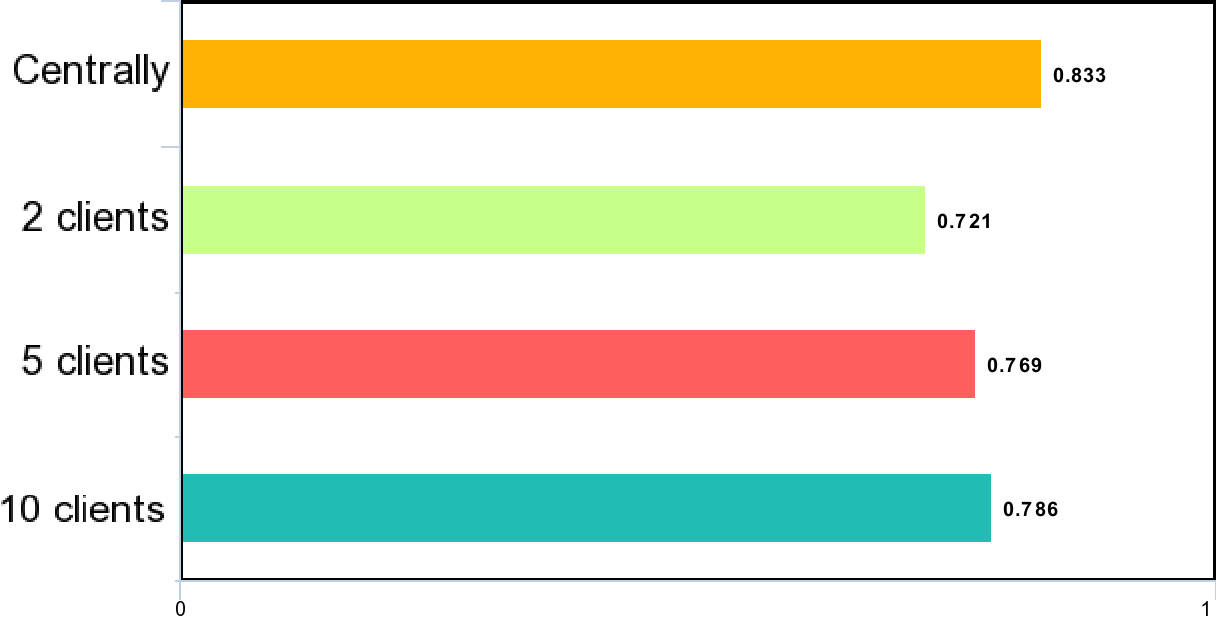}
    \caption{Effectiveness of MIA  in FL environment on CIFAR10 dataset for different number of clients.}
    \label{Tab1_fig}
\end{figure}

In this section, we present how successful membership inference attack is on ML models trained centrally or in a federated manner over $2$, $5$, and $10$ clients. \Cref{Tab1_fig} summarizes the accuracy of the MIA over the trained ML models.

% \begin{table}[!htb]
% 	\caption{Comparison of Membership Inference attack on Federally Learning} \label{Tab1}
% 	%\resizebox{\linewidth}{!}{%
% 		\begin{center}
% 			\begin{tabular}{ lllll}
% 				\hline
% 				Dataset    & Centrally & $2$ clients & $5$ clients & $10$ clients \\ \hline
% 				$CIFAR10$  & 0.833     & 0.721       & 0.769       & 0.786        \\
% 				% $CIFAR100$ &           & 0.765       & 0.765       & 0.796        \\ 
%     \hline
% 			\end{tabular}
% 		\end{center}
% 		%}
% \end{table}

When the model is trained centrally, we achieve almost $83.3\%$ attack accuracy. In contrast, for a federally trained model, we get a maximum accuracy of $78.6\%$ for ten clients. Therefore, when the model is trained federally, the attack is less effective.
 %We observe that the attack is less effective when the model is trained federally.  # This sentence is now redundant
 However, as the number of contributing clients increases this effect becomes less pronounced. Another observation, elaborated upon in the next section, training the model over federated architecture can mitigate the attacker's advantage gained by the proposed MIA-BAD approach. 

\begin{figure}[h!]
    \centering 
    \includegraphics[width=\linewidth]{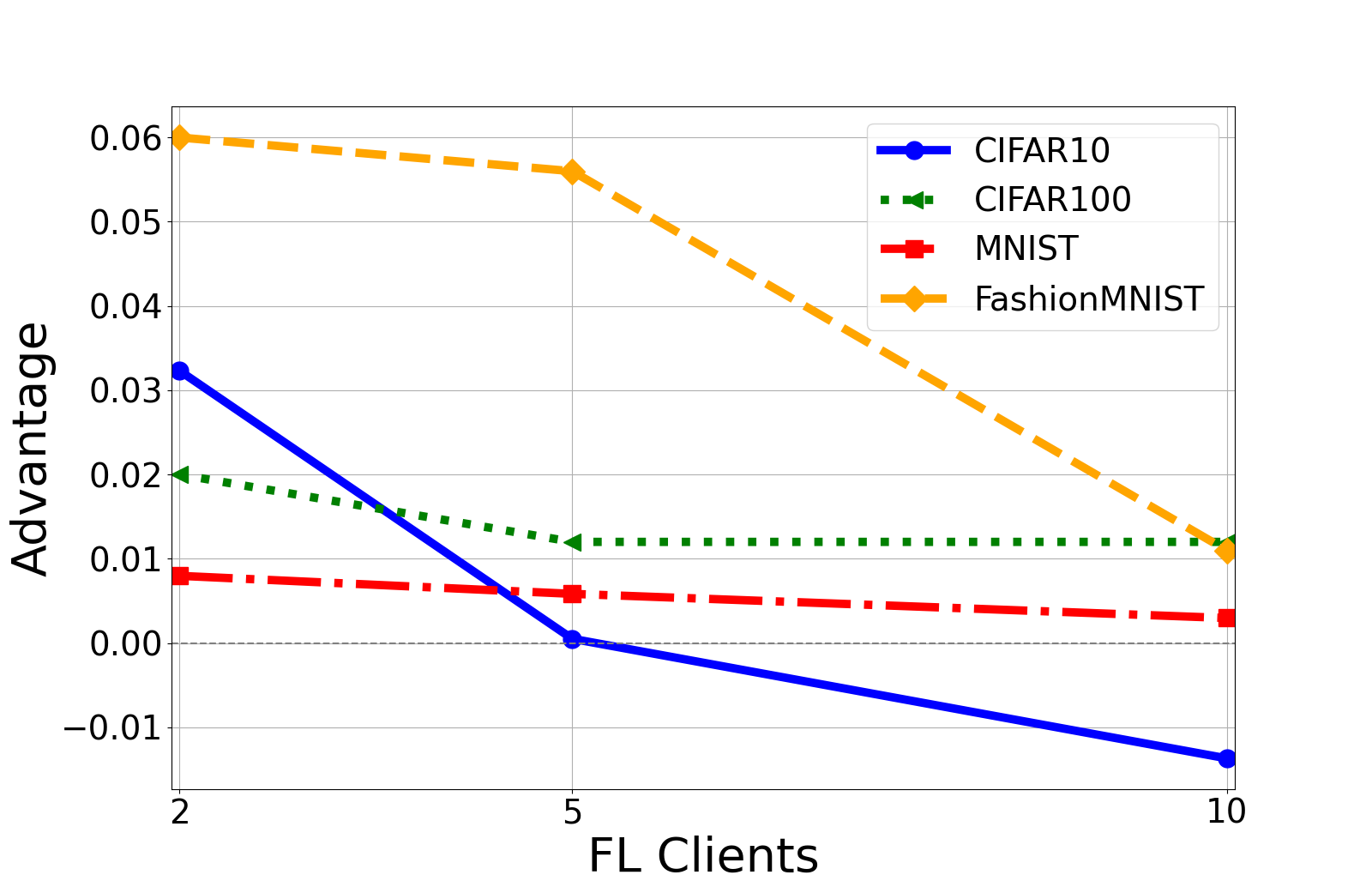}
    \caption{Comparison of the advantage of the MIA-BAD approach for 2, 5, and 10 clients as performed using CIFAR10, CIFAR100, MNIST, and FashionMNIST datasets.}
    \label{result}
\end{figure}

\subsection{Effect of Generating the Attack Dataset Batch-wise} \label{Res-C}
In this section, we highlight the experimental results of the proposed MIA-BAD approach. \Cref{Tab2} shows the effect of generating the attack dataset batch-wise when the target dataset is CIFAR10. From \cref{Tab2} we observe that the advantage of the MIA-BAD approach is not affected by the batch size. \Cref{Tab3} demonstrates the same result for MNIST, FashionMnist, and CIFAR100 respectively.

\begin{table*}[!htb]
	\caption{Comparison of accuracy of MIA-BAD on MNIST, FashionMNIST, and CIFAR100 datasets.} \label{Tab3}
    \centering
	% \resizebox{0.95\textwidth}{!}{%
    %\normalsize
		\begin{tabular}{ l|cc|cc|cc}
            \hline
            \multirow{2}{*}{Training Mode} & \multicolumn {2}{|c|}{MNIST} & \multicolumn {2}{c|}{FashionMNIST} & \multicolumn {2}{c}{CIFAR100} \\\cline{2-3}\cline{4-5}\cline{6-7}
                          & Sample      & Batch     & Sample      & Batch       & Sample      & Batch \\ \hline
            $2$ clients   & 0.838       & 0.846     & 0.747       & 0.807       & 0.701       & 0.721  \\
            $5$ clients   & 0.840       & 0.847     & 0.749       & 0.805       & 0.757       & 0.769  \\
            $10$ clients  & 0.849       & 0.852     & 0.787       & 0.798       & 0.786       & 0.798 \\ \hline
        \end{tabular}
	% }
\end{table*}

\Cref{Tab2} demonstrates how batch-wise generation of the attack dataset improves the attacker's advantage. We also observe that this effect is mostly independent of the batch size itself. Therefore, we only consider batch size 32 for the remaining datasets. We hypothesize that the marginal gain over the greater ensemble effect is perfectly countered by the reduced (attack) dataset size.  

\subsection{Effect of FL on the Proposed Approach}

From \cref{Tab2,Tab3}, we observe that the attacker's advantage over the MIA-BAD approach can be strongly countered by federated training of the ML model. The attacker's advantage is determined by the difference between batch-wise and sample-wise accuracy. In the case of CIFAR10, we consider the average batch-wise accuracy, for the remaining we consider batch size $32$. Empirically, we observe the number of federated clients is inversely proportional to the effect of the MIA-BAD approach. \Cref{result} demonstrates how the attacker's advantage through MIA-BAD can be mitigated through the federated training of ML models. In FashionMNIST, we observe the greatest advantage with the MIA-BAD approach, but nevertheless, it is mitigated with a high client count. Furthermore, with $10$ clients we observed a negative advantage in the case  of CIFAR10, completely mitigating the MIA-BAD advantage.

% \begin{figure}[h!]
%     \centering 
%     \includegraphics[width=\linewidth]{Fig_5.png}
%     \caption{Attacker's advantage through MIA-BAD, and its mitigation}
%     \label{result}
% \end{figure}

\section{Conclusion and future Scope}\label{conclusion}
In this paper, we propose a membership inference attack approach, MIA-BAD. We demonstrate that batch-wise attack dataset generation can provide an advantage to the adversary. We also demonstrate how the FL paradigm can be utilized to mitigate this approach. The proposed phenomenon is novel and promising and merits further investigation. In future works, we plan to investigate how the observed trends generalize over different ML models and more diverse datasets. We also plan to investigate if the proposed approach can be utilized with more advanced variants of the membership attacks.

%\nocite{*}
%\bibliographystyle{plain}
\bibliographystyle{IEEEtran}
\bibliography{reference.bib}
\end{document}